\documentclass[aps,prb,superscriptaddress,twocolumn]{revtex4}
\usepackage{graphicx}
\usepackage{amssymb}
\usepackage{amsmath}

\newcommand{\br}{{\bf r}}

\newcommand{\HF}{{\mathrm{HF}}}
\newcommand{\TF}{{\mathrm{TF}}}
\newcommand{\vW}{{\mathrm{vW}}}

\begin{document}

\title{Equilibrium geometries of low-lying isomers of some Li clusters,
   within Hartree-Fock theory plus
   bond order or MP2 correlation corrections}

\author{A. Grassi}
\affiliation{Dipartimento di Scienze Chimiche, Facolt\`a di Farmacia,
   Universit\`a di Catania,\\
Viale A. Doria, 6, I-95126 Catania, Italy}
\author{G. M. Lombardo}
\affiliation{Dipartimento di Scienze Chimiche, Facolt\`a di Farmacia,
   Universit\`a di Catania,\\
Viale A. Doria, 6, I-95126 Catania, Italy}
\author{G. G. N. Angilella}
\affiliation{Dipartimento di Fisica e Astronomia, Universit\`a di
   Catania,\\ and Istituto Nazionale per la Fisica della Materia,
   UdR di Catania,\\ Via S. Sofia, 64, I-95123 Catania, Italy}
\author{N. H. March}
\affiliation{Department of Physics, University of Antwerp (RUCA),\\
Groenenborgerlaan 171, B-2020 Antwerp, Belgium}
\affiliation{Oxford University, Oxford, England} 
\author{R. Pucci}
\affiliation{Dipartimento di Fisica e Astronomia, Universit\`a di
   Catania,\\ and Istituto Nazionale per la Fisica della Materia,
   UdR di Catania,\\ Via S. Sofia, 64, I-95123 Catania, Italy}

\date{\today}

\begin{abstract}
\medskip
In a recent study by Kornath \emph{et al.} [J. Chem. Phys. {\bf 118},
   6957 (2003)], the Li$_n$ clusters with $n=2,4$ and $8$ have been
   isolated in argon matrices at 15~K and characterized by Raman
   spectroscopy.
This has prompted us to carry out a theoretical study on such clusters
   up to $n=10$, using Hartree-Fock theory, plus low-order
   M\o{}ller-Plesset perturbation corrections.
To check against the above study of Kornath \emph{et al.,} as a
   by-product we have made the same approximations for $n=6$ and $8$
   as we have for $n=10$.
This has led us to emphasize trends with $n$ through the Li$_n$
   clusters for (i) ground-state energy, (ii) HOMO-LUMO energy gap,
   (iii) dissociation energy, and (iv) Hartree-Fock eigenvalue sum.
The role of electron correlation in distinguishing between low-lying
   isomers is plainly crucial, and will need a combination of
   experiment and theory to obtain decisive results such as that of
   Kornath \emph{et al.} for Li$_8$.
In particular, it is shown that Hartree-Fock theory plus bond order
   correlations does account for the experimentally observed symmetry
   T$_d$ symmetry for Li$_8$.
\end{abstract} 

\maketitle

\section{Background and outline}

Lithium clusters by now have been studied by a variety of experimental
   techniques which include electron spin resonance, laser induced
   fluorescence, depletion spectroscopy, photoionization, and Raman
   spectroscopy \cite{Kornath:03}.
This Ref.~\onlinecite{Kornath:03} has investigated, in particular, the
   lithium clusters Li$_n$, for $n=2,4$ and $8$.
This has been done by isolating these clusters in argon matrices at
   15~K.
Whereas Kornath \emph{et al.} \cite{Kornath:03} point out that most of
   the techniques listed above are difficult to correlate with the
   cluster geometries, their experimental work was able to determine
   the geometry of Li$_4$ as a rhombic structure (D$_{2h}$) and for
   Li$_8$ a hypertetrahedral structure (T$_d$) was shown to be in
   agreement with their Raman studies.

This experimental work has motivated the present theoretical study of
   Li$_n$ clusters.
Most attention is focussed here on $n=10$, but we have also
   included $n=6$ and $n=8$.
In the latter case, we must mention the previous theoretical work of
   Bonacic-Koutecky, Fantucci and Koutecky \cite{Bonacic-Koutecky:88}.

The outline of the paper is then as follows.
In Section~\ref{sec:1-10} some general trends of the electronic
   structure of Li$_n$ clusters are depicted with $n$ ranging from $1$
   to $10$.
We note that Kornath \emph{et al.} \cite{Kornath:03} studied $n=2$,
   $4$, and $8$ extensively using high quality techniques.
For Li$_8$ our more modest techniques essentially confirmed
   the findings of Kornath \emph{et 
   al.} \cite{Kornath:03} for the equilibrium geometry and the
   vibrational frequencies.
So we focus in Section~\ref{sec:6-8} dominantly on Li$_6$.
Section~\ref{sec:10} presents results for Li$_{10}$.
Finally, discussion of some additional trends, plus some suggestions
   for future work are covered in Section~\ref{sec:summary}.

\section{Quantum chemically predicted trends in L\lowercase{i$_n$}
   clusters for \lowercase{$n=1$} to $10$}
\label{sec:1-10}

In this Section we present our results of the calculations made on
   Li$_n$ clusters ($n=2,4,6,8,10$), by using the Gaussian package
   (G03 Linux version) \cite{Frisch:03}.
All the calculations were performed using the standard self-consistent
   field Hartree-Fock (HF) theory with the 6-311G$^\ast$ basis set.
The correlation energy was obtained using low-order M\o{}ller-Plesset (MP2)
   corrections \cite{Moeller:34}, considering all (valence and core)
   electrons.
For both HF and HF+MP2 level, full geometry optimization and
   vibrational analysis were performed for all Li clusters.

While Sections~\ref{sec:6-8} and \ref{sec:10} will develop further the
   systematic studies we have made especially for $n=6$ and $10$, with
   the approximate quantum chemical approach set out above, this
   Section will present the general trends predicted by HF+MP2 theory
   for $n=1$ to $10$.

\begin{figure}[t]
\centering
\includegraphics[width=\columnwidth]{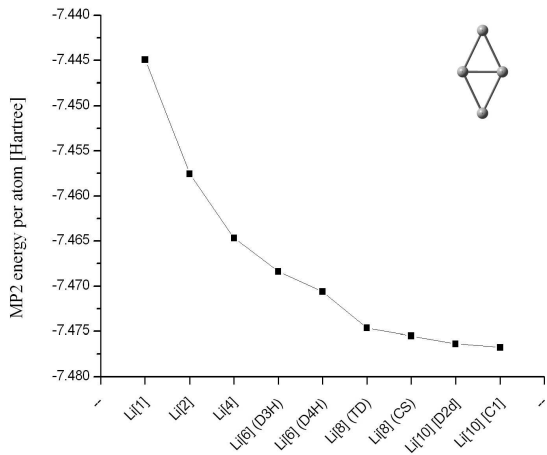}
\caption{Ground-state energy per atom of Li$_n$ clusters, with
   $n=1-10$.
\underline{\sl Inset:} accepted D$_{2h}$ planar structure of Li$_4$.
}
\label{fig:GS}
\end{figure}

To this end, Fig.~\ref{fig:GS} depicts the ground-state energy per
   atom for increasing $n$ from $1$ to $10$.
No commentary is needed on the results for $n=1-4$, except to show as
   an inset the accepted planar structure of Li$_4$.
The HF energy of this cluster is $-29.759$ and the lowering by
   second-order M\o{}ller-Plesset perturbation theory yields $-29.859$,
   both in Hartrees (as all energies below, unless stated otherwise).

As seen in Fig.~\ref{fig:GS}, two structures were found to be
   low-lying isomers of Li$_6$, with symmetries D$_{4h}$ and
   D$_{3h}$, and these will be discussed in detail in
   Section~\ref{sec:6-8} below.
Likewise, for Li$_8$, C$_s$ and T$_d$ structures emerge and again
   Section~\ref{sec:6-8} presents details.
Li$_{10}$ has the lowest MP2 energy per atom of the clusters
   considered and is the subject of Section~\ref{sec:10}.

\begin{figure}[b]
\centering
\includegraphics[width=\columnwidth]{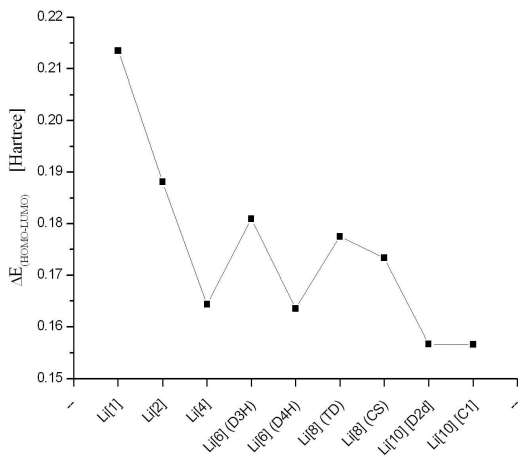}
\caption{LUMO-HOMO gap for Li$_n$ clusters as a function of $n$.}
\label{fig:HL}
\end{figure}

Again concerning general trends, Fig.~\ref{fig:HL} shows the variation
   of the HOMO-LUMO gap for Li$_n$ clusters with the number of atoms
   $n$ in the lithium cluster.
It is of interest to note that though the variation is irregular for
   $n=4-8$, the smallest gap is at $n=10$.
It is relevant to add here that solid lithium is metallic, with
   therefore zero energy gap.
Of course the gap at $n=10$ is still $\sim 5$~eV!
It is the trend, not the absolute numbers, which should be the focus
   here.

With this relatively brief discussion of trends we turn to consider
   our detailed results for Li$_6$ and, quite briefly, also Li$_8$ in
   Section~\ref{sec:6-8} immediately below.

\section{Geometry, energy, and vibrational frequencies of
L\lowercase{i}$_6$ and L\lowercase{i}$_8$ using HF+MP2 theory}
\label{sec:6-8}

\begin{figure}[t]
\centering
\includegraphics[width=0.49\columnwidth]{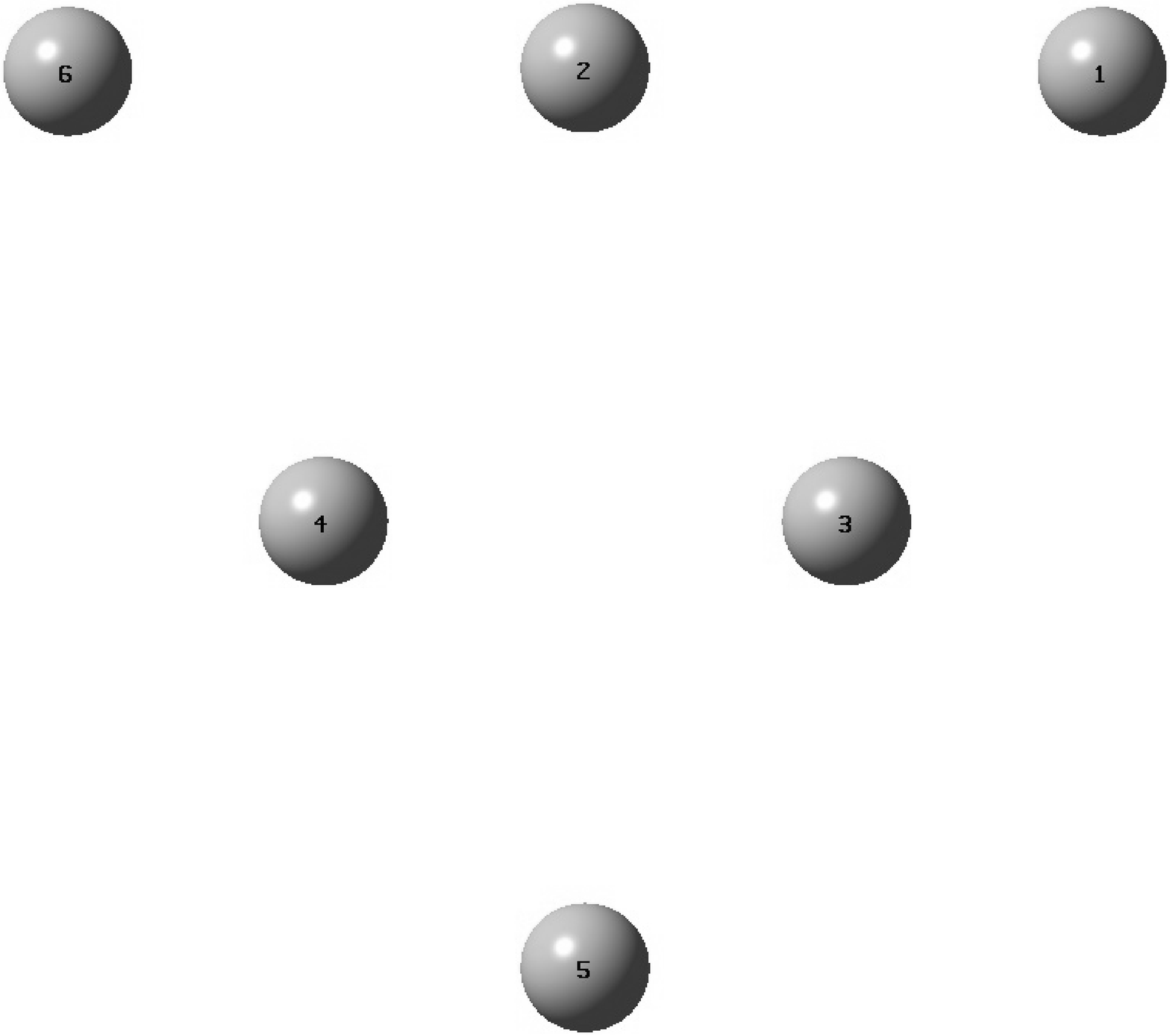}
\includegraphics[width=0.49\columnwidth]{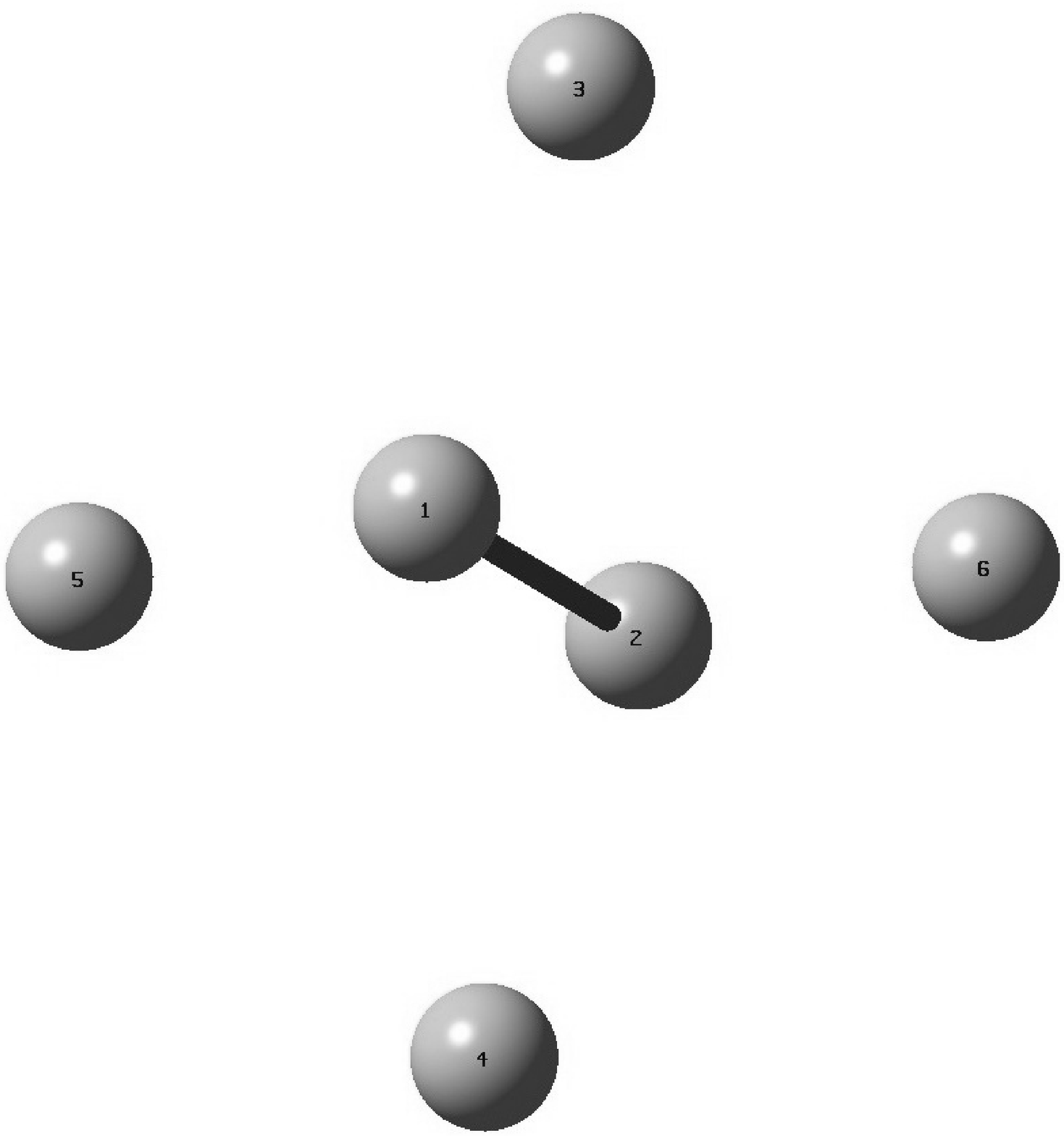}
\caption{Low-lying isomers of Li$_6$.
\underline{\sl Left:} Planar structure, with D$_{3h}$ symmetry.
\underline{\sl Right:} Most stable isomer predicted by HF+MP2, with
   D$_{4h}$ symmetry, and 
   a dimer-like Li$_2$
   structure perpendicular to the plane.
}
\label{fig:Li6}
\end{figure}

In Ref.~\onlinecite{Kornath:03}, a comment was made in their experimental
   considerations of a result which `may indirectly support the
   absence of a Li$_6$ species.'
Motivated by this, we have, by geometry optimization, `converged' on
   the geometries of two low-lying isomers which are depicted in
   Fig.~\ref{fig:Li6}.
The first of these is the planar structure (Fig.~\ref{fig:Li6}, left),
   with HF energy $-44.658$ and with addition of MP2 is $-44.810$.
The present HF+MP2 approach, however, predicts as the isomer we
   identify as the ground state, the planar square with a `diatomic'
   Li$_2$-like species perpendicular to the plane, through the center
   of gravity.
While this has a (very) slightly higher HF energy than the planar
   structure, it lies below that structure energetically when
   `correlation' treated at the MP2 level is added, the results being
   $E_\HF = -44.655$ and with the MP2 correction $-44.824$.

Turning quite briefly to Li$_8$, the geometry was settled as T$_d$
   from the Raman studies of Ref.~\onlinecite{Kornath:03}, whereas as can be
   seen from Fig.~\ref{fig:GS}, the C$_s$ symmetry is predicted as
   very slightly lower by our present approximation.
This is enough to illustrate the stringent test of many-electron
   approximate theories that is afforded by attempts to predict
   decisively the ground-state energies of isomers of Li$_n$ clusters.
However, the vibrational frequencies of Li$_8$ (T$_d$) are more in
   line with the observed values than those of the C$_s$ structure,
   which is somewhat encouraging in the light of the known geometry
   (T$_d$) of the lowest lying isomer.

We turn next to discuss, with fuller details, the Li$_{10}$ cluster.

\begin{figure}[t]
\centering
\includegraphics[width=0.49\columnwidth]{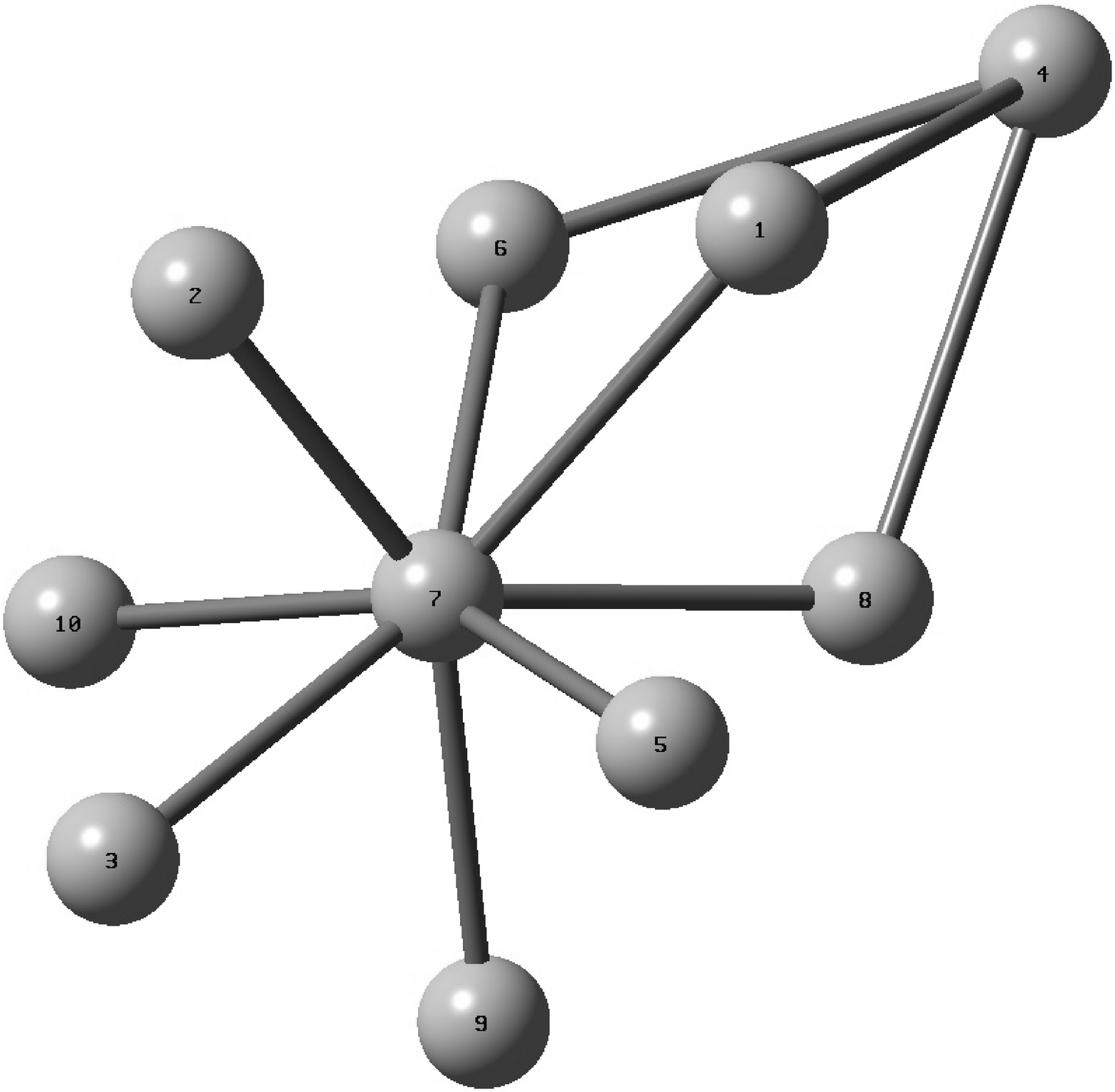}
\includegraphics[width=0.49\columnwidth]{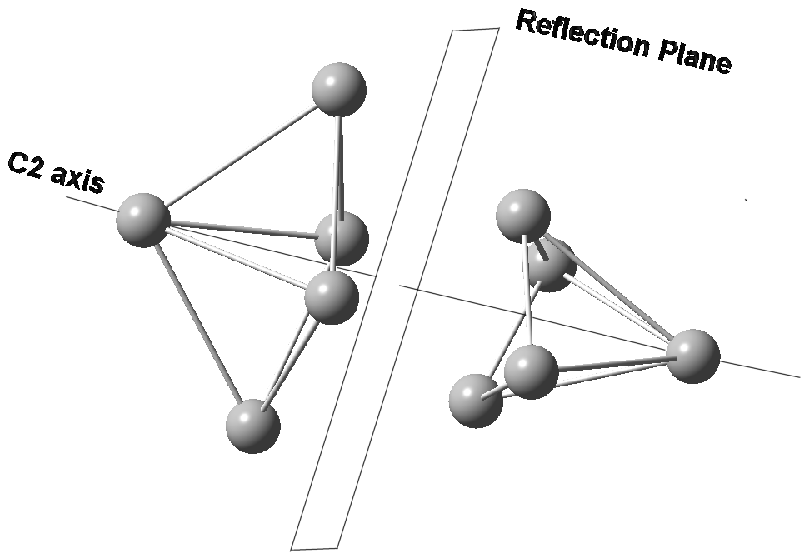}
\caption{Low-lying isomers of Li$_{10}$.
\underline{\sl Left:} Asymmetric isomer.
\underline{\sl Right:} Symmetric (D$_{2d}$) isomer.
The D$_{2d}$ isomer of Li$_{10}$ is characterized by
   reflection symmetry with respect to the plane indicated in the
   figure, and by 180$^\circ$ rotation symmetry with respect to the
   C$_2$ axis.
See also \protect\cite{epaps:Li10} for calculated properties of both
isomers.}
\label{fig:Li10}
\end{figure}

\section{Geometry, energy, and vibrational frequencies of two low-lying
   isomers of L\lowercase{i}$_{10}$}
\label{sec:10}

We have carried out similar calculations which have led us to two
   geometries of Li$_{10}$ which appear to be serious candidates for
   low-lying isomers.
The first of these is the very asymmetrical structure shown in
   Fig.~\ref{fig:Li10} (left).
Here, with an atom, roughly speaking, at the `center' of the cluster,
   there can, of course, be no symmetrical arrangement around it.

We have evaluated the total energies, the
   distance matrix (in \AA) and the calculated frequencies (in
   cm$^{-1}$) of the geometrical structure shown in 
   Fig.~\ref{fig:Li10} (left) \cite{epaps:Li10}.
The second structure we found as a low-lying isomer is shown in
   Fig.~\ref{fig:Li10} (right) \cite{epaps:Li10}.
This second isomer is characterized by D$_{2d}$ symmetry, and may be
   thought as obtained from the Li$_8$ cluster, by `adding' two more
   Li atoms (with, of course, different bond lengths and angles),
   shown in Fig.~\ref{fig:Li10} (right) in the top-right-hand and 
   bottom-left-hand positions.

\begin{figure}[t]
\centering
\includegraphics[width=\columnwidth]{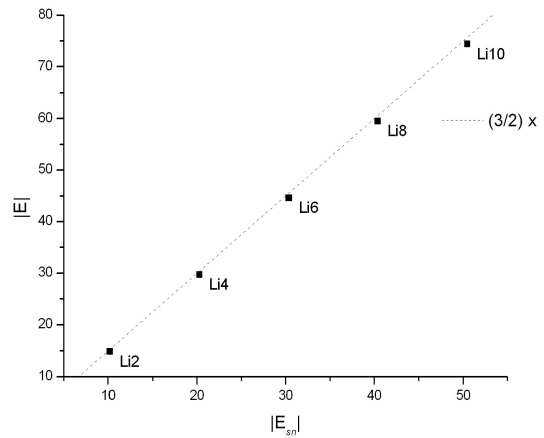}
\caption{Total energies $E_n$ versus eigenvalue sum $E_{sn}$ of Li$_n$
   ($n=2-10$).
The straight line is Eq.~(\protect\ref{eq:MP})
   \protect\cite{March:56}.
}
\label{fig:En}
\end{figure}

\section{Discussion and some future directions}
\label{sec:summary}

Let us begin this discussion of the results presented above by dealing
   first with further trends, to add to those reported in
   Section~\ref{sec:1-10}.

\subsection{Sum of eigenvalues ($E_{sn}$) related to total energy
   $E_n$, for $n$ atom cluster}

In Fig.~\ref{fig:En} we have plotted for $n=2$ to $10$ the total
   energies $E_n$ versus the eigenvalue sum $E_{sn}$ from the HF
   occupied eigenvalues $\epsilon_{in}$, defined by
\begin{equation}
E_{sn} = \sum_{\mathrm{occupied}~i} \epsilon_{in} .
\end{equation}

We find an approximately linear relation, which is compared with a
   theoretical prediction from initially Thomas-Fermi statistical
   theory, given by March and Plaskett \cite{March:56} for neutral
   atoms:
\begin{equation}
E = \frac{3}{2} E_s .
\label{eq:MP}
\end{equation}
The result (\ref{eq:MP}) was found, independently, by Ruedenberg
   \cite{Ruedenberg:77} from self-consistent field results for
   molecules, and its foundation in density functional theory was
   subsequently discussed by one of us \cite{March:77}.

\subsection{Trends in dissociation energy $D_n$ with number of atoms
   $n$ in Li$_n$ clusters defined as $|E_n - n E_1 |$}

As a further trend through the clusters Li$_n$ with $n$ going from $2$
   to $10$, we have collected in Table~\ref{tab:dissoc} results from
   our HF+MP2 calculations for the dissociation energy $D_n$.

\begin{table}[b]
\centering
\begin{tabular}{ccccc}
\hline
$n$ & Symmetry & $D_n$ & $N$ & $D_n /N^2$ \\
\hline
 2 & & 0.02530  &  6 & 0.000702778 \\
 4 & & 0.07896 & 12 & 0.000548333 \\
 6 & D$_{3h}$ & 0.14086 & 18 & 0.000434753 \\
 6 & D$_{4h}$ & 0.15411 & 18 & 0.000475648 \\
 8 & T$_d$ & 0.23765 & 24 & 0.000412587 \\
 8 & C$_s$ & 0.24471 & 24 & 0.000424844 \\
10 & D$_{2d}$ & 0.31460 & 30 & 0.000349555 \\
10 & C$_1$ & 0.31843 & 30 & 0.000353811 \\
\hline
\end{tabular}
\caption{Dissociation energy $D_n = |E_n - nE_1 |$, where $E_n$ is
   energy of Li$_n$ and $E_1$ that of isolated Li atom.
The second column refers to symmetry, when two low-lying isomers have
   been treated, while the fourth column refers to the total number of
   electrons, $N$.
All energies are in Hartree.
}
\label{tab:dissoc}
\end{table}

Mucci and March \cite{Mucci:83}, in early work, stressed the merit of
   Teller's theorem \cite{Teller:62}, which states that
   molecules/clusters do not bind 
   in any wholly local density approximation (LDA).
To avoid confusion with current terminology, `wholly LDA' in Teller's
   theorem refers to also treating the kinetic energy $T$ by the
   Thomas-Fermi (TF) result
\begin{equation}
T_\TF = c_k \int [\rho(\br)]^{5/3} \, d\br , \quad
c_k = \frac{3h^2}{10m} \left( \frac{3}{8\pi} \right)^{2/3} ,
\end{equation}
where $\rho(\br)$ denotes the electron density of the
   molecule/cluster.
If $T$ denotes the correct (single-particle) kinetic energy at the
   equilibrium geometry, Mucci and March \cite{Mucci:83} pointed out that
   the difference between $T$ and $T_\TF$ was entirely due to electron
   density gradients, \emph{e.g.} $\nabla\rho$, $\nabla^2 \rho$, etc.

\begin{figure}[t]
\centering
\includegraphics[width=\columnwidth]{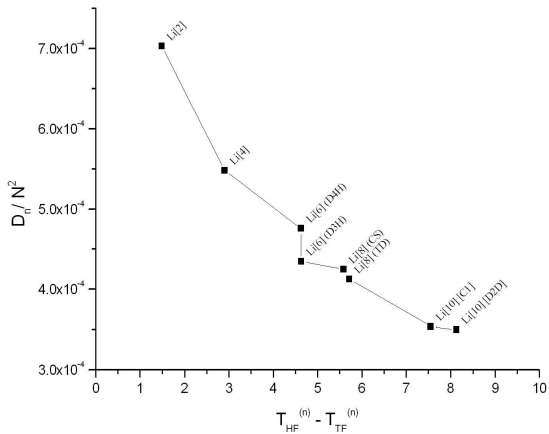}
\caption{Dissociation energy $D_n /N^2$, where $N$ is the total number
   of electrons in cluster for Li$_n$, versus HF kinetic energy
   $T_\HF^{(n)}$ minus Thomas-Fermi kinetic energy $T_\TF^{(n)}$.}
\label{fig:dissoc}
\end{figure}

This idea was followed up by Allan \emph{et al.} \cite{Allan:85}, who
   showed that $D/N^2$, with $N$ the total number of electrons in the
   molecule, indeed correlated with the simplest energy 
   constructed from $\nabla\rho$, namely the von~Weizs\"acker (vW)
   inhomogeneity kinetic energy defined by
\begin{equation}
T_\vW = \frac{1}{8} \int \frac{(\nabla\rho)^2}{\rho} \, d\br
   .
\end{equation}
This is the first-order correction due to density gradients in the
   difference $T-T_\TF$.
Therefore, in Fig.~\ref{fig:dissoc} we have plotted $D_n /N^2$,
   already recorded in Table~\ref{tab:dissoc}, versus $T_\HF^{(n)}
   -T_\TF^{(n)}$ 
   (where $T_\HF^{(n)} =-E_\HF^{(n)}$ at equilibrium from the virial
   theorem) for the series Li$_n$, with $n$ 
   again running from $2$ to $10$.

Subsequently, one of us \cite{March:91a} has given the inhomogeneous
   electron gas theory of molecular dissociation energies.
This reveals that $T_\vW^{1/6}$ is the crucial variable to
   use in characterizing $D_n /N^2$.
Therefore, in Fig.~\ref{fig:dissoc} we have made a further plot of
   $D_n /N^2$ already given in Table~\ref{tab:dissoc}, versus
   $|T_\HF^{(n)} - T_\TF^{(n)} |$, the latter quantity being expected
   to reflect $T_\vW^{(n)}$, albeit approximately.
There is a quite clear correlation from Fig.~\ref{fig:dissoc} between
   $D/N^2$ and the `gradient difference' $T_\HF - T_\TF$,
   substantiating the work reported in
   Refs.~\onlinecite{Mucci:83,Allan:85,March:91a}.

\subsection{Comments on individual clusters and especially role of
   electron correlation}

In cases when $n=6$ and $8$, two low-lying isomers have emerged from
   our studies using the HF+MP2 approximation.
A strong pointer to the importance of electron correlation in any
   quite decisive prediction of the true ground-state geometry is
   afforded by the two structures, considered for Li$_8$, where
   Kornath \emph{et al.} \cite{Kornath:03} have experimentally
   verified from their Raman studies that the correct structure is
   T$_d$.
This is indeed predicted to be lower in energy than C$_s$ in our HF
   studies, but the situation is (wrongly!) changed over by treating
   electron correlation at MP2 level.
However, as noted, HF+MP2 vibrational frequencies are considerably
   higher for T$_d$ than for C$_s$, the latter clearly being in marked
   disagreement with experiment!
On the other hand, a comparison of HF+MP2 energies and HF energy plus
   bond-order correlation energy (Sec.~\ref{ssec:correlation})
   shows that indeed inclusion of bond-order correlation allows to
   confirm the experimentally observed T$_d$ structure for Li$_8$.

Turning more briefly to Li$_6$, we find two low-lying isomers, both
   with lower energies than either 6 isolated Li atoms or 3 isolated
   Li$_2$ molecules.
Also stability with respect to isolated Li$_4$ and Li$_2$ components
   is clear.
Again, however, while D$_{4h}$ lies higher in energy than D$_{3h}$ at
   the HF level of approximation, including electron correlation at
   order MP2 reverses the ordering.
Our conclusion here then which seems to us firm is that two low lying
   isomers of Li$_6$ are found.
However, the energy ordering is not decisive, though our prejudice
   here is in favor of the MP2 addition, namely D$_{4h}$ symmetry, but
   that must remain conjecture until electron correlation is treated
   by more refined approaches such as the coupled cluster
   approximation.
Finally, the largest cluster studied here seems fairly `strongly'
   bound, being stable again with respect to 10 Li atoms, 5 Li$_2$
   dimers, and also the isolated `fragments' Li$_6$~$+$~Li$_4$ and
   Li$_8$~$+$~Li$_2$.
None of this must be taken to mean that our structure for Li$_{10}$ is
   the lowest-lying isomer of this cluster, though that is our
   prediction at the HF+MP2 level.

In the light of the importance of electron correlation, we shall now
   present some `heuristic' ideas on this subject, in which appeal
   will be made to bond-order versus bond-length relations which we
   have used earlier for polyatomic molecules \cite{Grassi:96}.
We note here that the study of Kornath \emph{et al.} \cite{Kornath:03}
   already includes extensive and careful \emph{ab initio} MO
   calculations with high level of treatment of electronic correlations.

\subsection{Some approximate considerations on the magnitude of electron
   correlation energy in the L\lowercase{i$_n$} clusters with
   \lowercase{$n$} from $2$ to $10$}
\label{ssec:correlation}

\begin{figure}[t]
\centering
\includegraphics[height=\columnwidth,angle=-90]{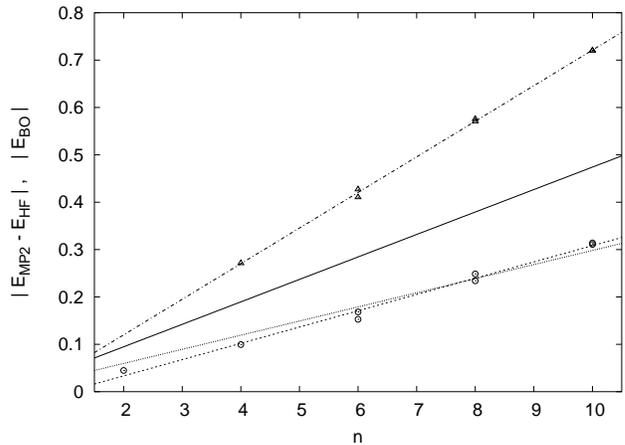}
\caption{Different estimates of the magnitude of the electron
   correlation energy in Li$_n$ clusters versus $n$
   (Sec.~\protect\ref{ssec:correlation}). 
Solid line is based on `estimate' of $n$ `independent' K shell
   contributions and here correlation energy is directly proportional
   to $n$.
Open circles show results due to MP2 perturbation theory.
Two different `linear' approximations are represented by the dashed
   lines.
Small-dashed line passes through origin at $n=0$.
Long-dashed line is a (slightly) better fit to the MP2 results shown
   as circles.
Open triangles show results due to bond-order correlation, for which
   the dash-dotted line is a guide for the eye.
For Li$_6$, Li$_8$ and Li$_{10}$, values for two low-lying isomers
   studied in each case are shown.
}
\label{fig:linear}
\end{figure}

Using the L\"owdin definition
   \cite{Loewdin:55a,Loewdin:55b,Loewdin:55c} of electron correlation,
   say $E_c^{\mathrm{L}}$, as
\begin{equation}
E_c^{\mathrm{L}(n)}= E_{\mathrm{exact}}^{(n)} - E_\HF^{(n)} ,
\end{equation}
we might use, as a first approximation to $E_c^{\mathrm{L}(n)}$ the
   correlation energy in the K shells, namely
\begin{equation}
E_c^{\mathrm{L}(n)} \approx n E_c^{\mathrm{L}(1)} .
\label{eq:L2}
\end{equation}
One can utilize, for example, the recent study of Alonso \emph{et al.}
   \cite{Alonso:03} to estimate $E_c^{\mathrm{L}(1)}$ for the Li atom
   ground state.
Their pairing energy $E_{ss}$ is, for neutral atoms, given in their
   Table~2 as $-1.28$~eV, and hence for the Li$_{10}$ cluster we
   estimate (essentially from 10 separate K shells) the correlation
   energy in magnitude to be $0.47$~Hartree.
The upper curve in Fig.~\ref{fig:linear} shows this K-shell--like
   magnitude of $E_c$ for the 
   Li$_n$ clusters under consideration.
Of course, the original $2s$ electrons will form molecular orbitals
   where again there is pairing of electrons with antiparallel spins,
   and we therefore suppose that the uppermost curve in
   Fig.~\ref{fig:linear} will be below the `true' magnitude of the
   correlation energy curve.

For comparison with this K shell estimate, we have plotted the
   `correlation energy' given by HF+MP2 perturbation theory.
The circles in Fig.~\ref{fig:linear} show our results for Li$_n$
   ($n=2,4,6,8$ and $10$).
Two low-lying isomers were considered in the present study of both
   Li$_6$ and Li$_8$.
The small-dashed curve is proportional to $n$ as in Eq.~(\ref{eq:L2}),
   whereas the long-dashed curve is linear in $n$ but does not pass
   through the origin, in contrast to Eq.~(\ref{eq:L2}).

Evidently, while MP2 corrections obviously improve the HF energies,
   they yield only $\sim 50$ to $60$~\% the Li correlation energies,
   the poorest result being for the dimer.

\begin{table}[b]
\centering
\begin{tabular}{cccccc}
\hline
$n$ & & $E_c^{\mathrm{BO}}$ & $E_c^{\mathrm{MP2}}$ &
   $E_c^{\mathrm{BO}} - E_c^{\mathrm{MP2}}$ & $\epsilon_c
   = E_c^{\mathrm{BO}} /Z$ \\
\hline
 4 &          & $-0.27126$ & $-0.09887$ & $-0.17239$ & $-0.06782$ \\
 6 & D$_{3h}$ & $-0.41049$ & $-0.15191$ & $-0.25858$ & $-0.06842$ \\
 6 & D$_{4h}$ & $-0.42696$ & $-0.16724$ & $-0.25972$ & $-0.07116$ \\
 8 & T$_d$    & $-0.57063$ & $-0.23078$ & $-0.33985$ & $-0.07133$ \\
 8 & C$_s$    & $-0.57561$ & $-0.24506$ & $-0.33055$ & $-0.07195$ \\
10 & D$_{2d}$ & $-0.72000$ & $-0.30339$ & $-0.41661$ & $-0.07200$ \\
10 & C$_1$    & $-0.72041$ & $-0.31088$ & $-0.40953$ & $-0.07204$ \\
\hline
\end{tabular}
\caption{Correlation energies from both bond order
   (BO) and MP2 calculations.
The second column lists the symmetry of the cluster.
All energies are in Hartree.
}
\label{tab:correlation}
\end{table}

Following the technique reported in Ref.~\onlinecite{Grassi:96}, we have
   calculated the correlation energies for lithium clusters using the
   HF bond orders.
Fig.~\ref{fig:linear} compares our results with the values obtained
   from MP2 calculations.
In particular, the experimental correlation energy
   $E_c^{\mathrm{exp}}$ is defined as:
\begin{equation}
E_c^{\mathrm{exp}} = E_S - E_\HF ,
\end{equation}
with $E_S$ the exact Schr\"odinger energy and $E_\HF$ the Hartree-Fock
   energy.
Following Ref.~\onlinecite{Grassi:96}, we calculate the theoretical
   correlation energy $E_c^{\mathrm{BO}}$ as the sum of atomic and
   bond contributions:
\begin{equation}
E_c^{\mathrm{BO}} = \sum_{i=1}^N E_S^A n_{\mathrm{eff}} +
   \sum_{i=1}^{N-1} \sum_{j=i+1}^N a_{ij} B_{ij} ,
\label{eq:corrBO}
\end{equation}
with $N$ the number of atoms in the cluster.
The $E_S^A$ term is the atomic Schr\"odinger correlation energy that
   can be obtained from experimental data, as explained in
   Ref.~\onlinecite{Grassi:96}, $n_{\mathrm{eff}}$ denotes the effective
   atomic electron number, which takes into account the electrons not
   involved in the bonds, \emph{i.e.}
\begin{equation}
n_{\mathrm{eff}} = \frac{Z-n_{\mathrm{bond}}}{Z} ,
\end{equation}
with $Z$ the atomic number and $n_{\mathrm{bond}}$ the total number of
   electrons involved in molecular bonds for each atom.
$B_{ij}$ is the bond order (BO) between atoms $i$ and $j$, as obtained
   from HF calculations, and $a_{ij}$ is a parameter depending on the
   particular bond $i-j$.
For Li clusters, this parameter was obtained from the experimental
   correlation energy of Li$_2$ molecule and its value is
   $0.06157$~a.u.
While up to now all binding and correlation energies have
   been calculated within HF+MP2, with full geometry optimization, the
   bond order correlation energies discussed here refer to the
   optimized geometries within HF.

\begin{table*}[t]
\begin{tabular}{ccccccccc}
\hline
$n$ & & $E_\HF$ & $E_{\mathrm{BO}}$ &
   $E_{\mathrm{MP2}}$ & & $\Delta E_\HF$
   &
   $\Delta E_{\mathrm{BO}}$ & $\Delta
   E_{\mathrm{MP2}}$ \\ 
& & & & & & ($\times 1000$) & ($\times 1000$) & ($\times 1000$)\\ 
\hline
 4 &          & $-29.75973$ & $-30.03099$ & $-29.85860$ &          &
   & & \\
 6 & D$_{3h}$ & $-44.65841$ & $-45.06890$ & $-44.81032$ &          &
   & & \\
 6 & D$_{4h}$ & $-44.65633$ & $-45.08329$ & $-44.82357$ &
   D$_{4h}$~$-$~D$_{3h}$& $2.0800$ &
   $-14.390$ & $-13.252$ \\
 8 & T$_d$    & $-59.56615$ & $-60.13678$ & $-59.79693$ &          &
   & & \\
 8 & C$_s$    & $-59.55893$ & $-60.13454$ & $-59.80399$ &
   C$_{s}$~$-$~T$_{d}$ & $7.2191$ 
   & $2.239$ &  $-7.058$ \\
10 & D$_{2d}$ & $-74.46031$ & $-75.18031$ & $-74.76370$ &          &
   & & \\
10 & C$_1$    & $-74.45665$ & $-75.17706$ & $-74.76753$ &
   C$_{1}$~$-$~D$_{2d}$ & $3.6587$ &
   $3.249$ & $-3.827$ \\ 
\hline
\end{tabular}
\caption{Hartree-Fock and total energies from both bond order
   (BO) and MP2 calculations, defined as $E_{\mathrm{BO}} = E_\HF +
   E_c^{\mathrm{BO}}$ and $E_{\mathrm{MP2}} = E_\HF +
   E_c^{\mathrm{MP2}}$.
The second column lists the symmetry of the cluster.
Last three columns list differences of these energies between clusters
   with same $n$ but different symmetry (D$_{4h}$~$-$~D$_{3h}$, for
   $n=6$; C$_{s}$~$-$~T$_{d}$, for
   $n=8$; C$_{1}$~$-$~D$_{2d}$, for
   $n=10$).
All energies are in Hartree.}
\label{tab:HFcorr}
\end{table*}

Table~\ref{tab:correlation} reports the calculated correlation
   energies for Li clusters obtained from MP2 calculations
   ($E_c^{\mathrm{MP2}}$) and by using Eq.~(\ref{eq:corrBO})
   ($E_c^{\mathrm{BO}}$), along with the HF energies.
As shown in Table~\ref{tab:correlation} and in Fig.~\ref{fig:linear},
   the calculated correlation energies from Eq.~(\ref{eq:corrBO}) are
   higher than the MP2 energies, giving for each Li cluster a total
   molecular energy less than that calculated at HF+MP2 level.
Moreover, it is noteworthy that the difference $E_c^{\mathrm{MP2}} -
   E_c^{\mathrm{BO}}$ increases with increasing number of atoms in the
   cluster, and that the correlation energy per electron, $\epsilon_c
   = E_c^{\mathrm{BO}} /Z$, is about constant for all clusters,
   indicating that the total correlation energy is proportional to the
   total number of electrons in the molecule.
In addition, a comparison of the total HF+MP2 and the HF+BO energies in
   Table~\ref{tab:HFcorr} shows
   that, while for $n=6$ both models predict the cluster with $D_{4h}$
   symmetry to be stabler than the cluster with $D_{3h}$ symmetry, at
   variance with the HF result,
   both for $n=8$ and $n=10$ only the HF+BO energies do predict the
   $T_d$ and $D_{2d}$ to be stabler than the $C_s$ and $C_1$,
   respectively, in agreement with the purely HF results and, in the
   case of Li$_8$, in agreement with the experimental result
   \cite{Kornath:03}.

Fig.~\ref{fig:linear} shows that the MP2 correlation energy estimated
   with the 6-311G$^\ast$ basis set is systematically much smaller
   than the sum of the $(1s)^2$ atomic pair correlation.
The referee has pointed out to us that this probably results from the
   basic set deficiency for the inner shell correlation.

\begin{acknowledgments}
N.H.M. brought his contribution to the present study to fruition
   during a visit to Catania in 2003.
He thanks the Department of Physics and Astronomy for the stimulating
   atmosphere and for generous hospitality.
\end{acknowledgments}

\medskip

\begin{small}
\bibliographystyle{apsrev}
\bibliography{a,b,c,d,e,f,g,h,i,j,k,l,m,n,o,p,q,r,s,t,u,v,w,x,y,z,zzproceedings,Angilella,notes}
\end{small}

\end{document}